\documentclass[aps,prc,letterpaper,11pt,twoside,tightenlines,nofootinbib,showpacs,preprint]{revtex4-1}
\usepackage{graphicx}
\usepackage[sort&compress]{natbib}
\usepackage{subfigure}
\usepackage{amsmath}
\usepackage{amsfonts}
\usepackage{cancel}

\begin{document}

\arraycolsep1.5pt

\newcommand{\Ima}{\textrm{Im}}
\newcommand{\Rea}{\textrm{Re}}
\newcommand{\mev}{\textrm{ MeV}}
\newcommand{\be}{\begin{equation}}
\newcommand{\ee}{\end{equation}}
\newcommand{\ba}{\begin{eqnarray}}
\newcommand{\ea}{\end{eqnarray}}
\newcommand{\gev}{\textrm{ GeV}}
\newcommand{\nn}{{\nonumber}}
\newcommand{\dtres}{d^{\hspace{0.1mm} 3}\hspace{-0.5mm}}

\title{Experimental constraints on the $\omega$-nucleus real potential}
\author{~S.~Friedrich$^{1}$,~K.~Makonyi$^{1,a}$,~V.~Metag$^{1,5}$,~D.~Bayadilov$^{2,3}$,~B.~Bantes$^{4}$,~R.~Beck$^{2}$,\\~Y.~A.~Beloglazov$^{2,3}$,~S.~B\"ose$^{2}$,~K.-T.~Brinkmann$^{1}$,~Th.~Challand$^{4}$,~V.~Crede$^{5}$,~T.~Dahlke$^2$,\\~F.~Dietz$^{1}$,~P.~Drexler$^{1}$,~H.~Eberhardt$^{6}$,~D.~Elsner$^{6}$,~R.~Ewald$^{6}$,~K.~Fornet-Ponse$^{6}$,\\~F.~Frommberger$^6$,~Ch.~Funke$^{2}$,~M.~Gottschall$^2$,~A.~Gridnev$^{2,3}$,~M.~Gr\"uner$^2$,~E.~Gutz$^{1,2}$,\\~Ch.~Hammann$^2$,~D.~Hammann$^{6}$,
~J.~Hannappel$^6$,~J.~Hartmann$^{2}$,~W.~Hillert$^{6}$,~S.~Hirenzaki$^{7}$,\\~P.~Hoffmeister$^2$,~Ch.~Honisch$^2$,~I.~Jaegle$^{4,b}$, 
~D.~Kaiser$^2$,~H.~Kalinowsky$^2$,~S.~Kammer$^6$,\\~I.~Keshelashvili$^4$,~V.~Kleber$^6$,~F.~Klein$^6$,~B.~Krusche$^4$,~M.~Lang$^2$,~I.~V.~Lopatin$^{2,3}$,
\\~Y.~Maghrbi$^{4,c}$,~M.~Nanova$^{1}$,~H.~Nagahiro$^{7}$,~J.~M\"uller$^2$,~T.~Odenthal$^2$,~D.~Piontek$^{2}$,\\~T.~Rostomyan$^{4}$,~S.~Schaepe$^2$,~Ch.~Schmidt$^2$,~H.~Schmieden$^6$,~R.~Schmitz$^2$,~T.~Seifen$^2$,\\~A.~Thiel$^2$,~U.~Thoma$^2$,~H.~van~Pee$^2$,~D.~Walther$^2$,~J.~Weil$^{8,d}$,~Ch..~Wendel$^2$,\\U.~Wiedner$^{9}$,~A.~Wilson$^{2,5}$,~A.~Winnebeck$^2$, and F.~Zenke$^2$\\
(The CBELSA/TAPS Collaboration)}
\affiliation {
{$^{1}$II. Physikalisches Institut, Universit\"at Gie{\ss}en, Germany}\\
{$^{2}$Helmholtz-Institut f\"ur Strahlen- u. Kernphysik, Universit\"at Bonn, Germany}\\
{$^{3}$Petersburg Nuclear Physics Institute, Gatchina, Russia}\\
{$^{4}$Departement Physik, Universit\"at Basel, Switzerland}\\
{$^{5}$Department of Physics, Florida State University, Tallahassee, FL, USA}\\
{$^{6}$Physikalisches Institut, Universit\"at Bonn, Germany}\\
{$^{7}$Department of Physics, Nara Women's University, Nara 630-8506, Japan}\\
{$^{8}$Institut f\"ur Theoretische Physik, Universit\"at Gie{\ss}en, Germany}\\
{$^{9}$Physikalisches Institut, Universit\"at Bochum, Germany}\\
{$^{a}$Current address: University of Stockholm, Sweden}\\
{$^{b}$Current address: Hawaii University, USA} \\
{$^{c}$Current address: Texas A$\&$M University, Qatar}\\
{$^{d}$Current address: Institut f\"ur Theoretische Physik, Universit\"at Frankfurt, Germany}\\
}
\date{\today}
\begin{abstract} 
In a search for $\omega$ mesic states, the production of $\omega$-mesons in coincidence with forward going protons has been studied in photon induced reactions on $^{12}$C for incident photon energies of 1250 - 3100 MeV. The $\pi^0 \gamma$ pairs from decays of bound or quasi-free $\omega$-mesons have been measured with the CBELSA/TAPS detector system in coincidence with protons registered in the MiniTAPS forward array. Structures in the total energy distribution of the $\pi^0 \gamma$ pairs, which would indicate the population and decay of bound $\omega~^{11}$B states, are not observed. The $\pi^0 \gamma$ cross section of 0.3 nb/MeV/sr observed in the bound state energy regime between -100 and 0 MeV may be accounted for by yield leaking into the bound state regime because of the large in-medium width of the $\omega$-meson.  A comparison of the measured total energy distribution with calculations suggests the real part $V_0$ of the  $\omega~^{11}$B potential to be small and only weakly attractive with $V_0(\rho=\rho_0) = -15\pm$ 35(stat) $\pm$20(syst) MeV in contrast to some theoretical predictions of attractive potentials with a depth of 100 - 150 MeV.
 \end{abstract}
\maketitle

\section{Introduction}
\label{Intro}
Mesons and baryons are the relevant degrees of freedom in strong interaction physics in the few GeV energy regime. The study of the interaction of mesons with nuclei has provided important information on the strong force in this energy range. While the interaction of long lived charged mesons like $\pi^{\pm}$ or K$^{\pm}$ with nuclei can be studied with secondary meson beams, the majority of short-lived mesons can only be investigated by producing them on a nucleus and using the same nucleus as a laboratory for studying their interaction with the medium. A question of particular interest is whether this interaction is attractive and strong enough to form bound states of mesons and nuclei. 

The existence of deeply bound $\pi^-$-nucleus  states has been established in a series of measurements at the fragment separator FRS at GSI \cite{Gilg,Itahashi, Geissel,Suzuki} by studying recoil-free production of $\pi^-$-mesons in the (d,${}^3\textrm{He})$ reaction on various Pb and Sn isotopes. The widths of these states were found to be about 1/10 of the binding energy which is essential for their experimental identification. The origin of these states is due to a superposition of the attractive Coulomb force and the repulsive s-wave $\pi^-$-nucleus interaction which gives rise to a potential pocket, leading to a halo-like $\pi^-$ distribution around the nucleus \cite{Kienle_Yamazaki}. 

It is of utmost importance for our understanding of the strong interaction whether neutral mesons which are only subject to the strong interaction can also form meson-nucleus bound states. Theoretical predictions for $\eta$ \cite{Garcia_Recio,Nagahiro_eta,Bass,Friedman,Cieply}, $\omega$  \cite{Marco_Weise,Kaskulov,Nagahiro_omega}, and $ \eta^\prime$ \cite{Nagahiro_Zaki,Nagahiro,Jido,Nagahiro_etaprime} mesic states have been made. In case of the $\eta$-meson the predicted widths are comparable to or even larger than the binding energies. This implies the strength of these states partially extends into the continuum, allowing for free meson emission. Experimental indications for surprisingly large $\eta$ yields in coherent $\eta$ photo production near threshold have been reported \cite{Pfeiffer,Pheron}. Similar observations have been made by the COSY-ANKE collaboration by studying the $\textrm{pd} \rightarrow \eta {}^{3}\textrm{He}$ reaction \cite{Goslawski}. In both cases, the very strong rise of the cross section near threshold is taken to be indicative of a quasi-bound state close to threshold. A direct observation of an $\eta$ bound state has been claimed by the COSY-GEM collaboration in the $\textrm{p}+{}^{27}\textrm{Al} \rightarrow {}^{3}\textrm{He} +{}_{\eta}^{25}\textrm{Mg} \rightarrow {}^3\textrm{He} + \textrm{p} +\pi^- +\textrm{X}$ reaction \cite{Budzanowski}. For the d+d$\rightarrow {}_{\eta}^4\textrm{He}\rightarrow{}^3\textrm{He} + \textrm{p} +\pi^- $ reaction, only an upper limit for the population of an $\eta$ bound state was reported by the WASA@COSY collaboration \cite{WASA}.

The case for $\eta^\prime$ mesic states appears to be promising experimentally because of the relatively narrow in-medium width of the $\eta^\prime$-meson of about 20 MeV at normal nuclear matter density as determined in a transparency ratio measurement \cite{Nanova}. Corresponding experiments are planned at the GSI fragment separator and later at the SuperFRS at FAIR \cite{Kenta,Nagahiro_Kenta} as well as at the BGO-OD spectrometer at the electron accelerator ELSA \cite{Volker}.

In this work we report on a search for $\omega$ mesic states in a photo nuclear reaction exploiting near recoil-free $\omega$ production. In a  ($\gamma,\textrm{p} \omega$) reaction on a nuclear target, the forward going proton takes over most of the momentum of the incoming photon beam, leaving the $\omega$-meson almost at rest so it can be captured by the nucleus in case of an attractive $\omega$-nucleus interaction. For a nucleon at rest, recoilless production occurs at an incident photon energy of 2.7 GeV, but - as discussed in \cite{Kaskulov} -momentum transfers to the $\omega$-meson are less than about 400 MeV/$c$ within the whole energy range between 1250 - 3100 MeV as long as the outgoing proton is confined to laboratory polar angles of $1^{\circ}-11^{\circ}$. It is important to exploit the full incident photon energy range as the cross section for slow $\omega$-mesons, going backwards in the center-of-mass system, is about 10 times larger at $E_{\gamma}$ = 1.2 GeV than at 2.7 GeV \cite{SAPHIR}. In the search for $\omega$ mesic states one has to be aware of the large in-medium width of the $\omega$-meson, determined to be $\Gamma$ = 140 MeV at normal nuclear matter density for recoil momenta of 1.1 GeV/$c$ \cite{Kotulla}. This may make it difficult to observe distinct structures in the cross section, indicating the formation of an $\omega$ mesic state, even though the $\omega$ momenta and thus the $\omega$ width in the present experiment are expected to be much smaller than in  \cite{Kotulla}.

\section{Experiment and data analysis}

The experiment was performed at the electron stretcher accelerator ELSA in Bonn, using the combined Crystal Barrel (CB) (1320 CsI modules) \cite{Aker} and MiniTAPS detectors (216 BaF$_2$ modules) \cite{Gabler} subtending polar angles of 11$^{\circ} - 156^{\circ}$ and $1^{\circ}-11^{\circ}$, respectively, and the full azimuthal angular range, thereby covering 96$\%$ of the full solid angle. Tagged photons with energies of 0.65 - 3.1 GeV, produced via bremsstrahlung at a rate of 6 - 10 MHz, impinged on a 15 mm thick ${}^{12}\textrm{C}$ target, corresponding to 6$\%$ of a radiation length $X_0$. As a reference measurement, a LH$_2$ target of 5.1 cm length (0.6$\%$~of~$ X_0$) was irradiated under the same experimental conditions. The photon flux through the target was determined by counting the photons reaching the $\gamma$ intensity detector at the end of the setup in coincidence with electrons registered in the tagging system. The data were collected during two running periods of 830~h and 450~h for the C and LH$_2$ target, respectively. Events with $\omega $ candidates were selected with suitable multiplicity trigger conditions, requiring at least one hit in MiniTAPS and more than two hits in the CB, derived from a fast cluster recognition encoder. A more detailed description of the detector setup and the running conditions can be found in \cite{AThiel}. 
The $\omega$-mesons were identified via the $\omega \rightarrow \pi^{0} \gamma$ decay channel, which has a branching ratio of 8.3\% \cite{PDG}. For the reconstruction of the $\omega$-proton 
\begin{figure}[!ht]
 \resizebox{0.5\textwidth}{!}{
    \includegraphics[height=0.4\textheight]{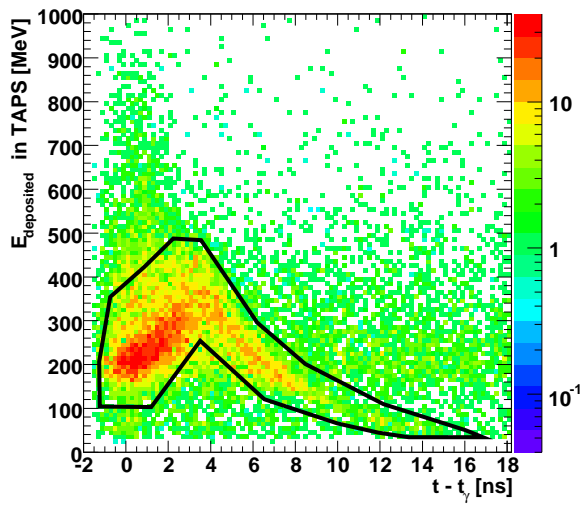}}
\caption{Measured energy deposited in the BaF$_2$ modules of the MiniTAPS forward array versus the time-of-flight of charged particles relative to photons for events with one charged hit in MiniTAPS and exactly 3 neutral hits in CB. The cut selecting protons is indicated. Protons with kinetic energies larger than 400 MeV (flight times t-t$_{\gamma} \le $3.5 ns) punch through the BaF$_2$ modules and deposit only part of their energy. The data were taken with the carbon target.}
\label{fig:p_band}
\end{figure}
pairs, only events with exactly 3 neutral hits and 1 charged particle hit in MiniTAPS were selected. Protons were identified in the BaF$_2$ modules of MiniTAPS by exploiting the characteristic correlation between deposited energy and time-of-flight (see Fig.~\ref{fig:p_band}). In addition, an aerogel Cherenkov detector with an index of refraction of n=1.05 was used to veto electrons and charged pions in the angular range covered by MiniTAPS. Events of interest were selected and the background suppressed by several kinematical cuts. Only events with incident photon energies larger than 1250 MeV were processed. A $\pm 3 \sigma $ cut on the missing mass, calculated for the $\pi^0 \gamma$ pair and the entrance channel, was applied. Photons were required to have energies larger than 50 MeV (to suppress shower splittings) and to be in the polar angular range of 14$^{\circ}$ to 156$^{\circ}$ while protons had to be in the polar angular range 1$^{\circ}$ - 11$^{\circ}$. With the 3 photons per event the invariant mass of all photon pairs was calculated and the one combination closest to the $\pi^0$ mass of 135 MeV/$c^2$ was taken to be the $\pi^0$. The energy of the bachelor photon was then requested to be larger than 200 MeV to suppress background from $\pi^0 \pi^0$ and $\pi^0 \eta$ events. Events with rescattered $\pi^0$-mesons from $\omega \rightarrow \pi^0 \gamma$ decays within the nucleus were suppressed by requesting the kinetic energy of the $\pi^0$ to be larger than 120 MeV \cite{Messchendorp,Kaskulov}.  
\begin{figure}[!ht]
 \resizebox{0.8\textwidth}{!}{
    \includegraphics[height=0.89\textheight,angle=90]{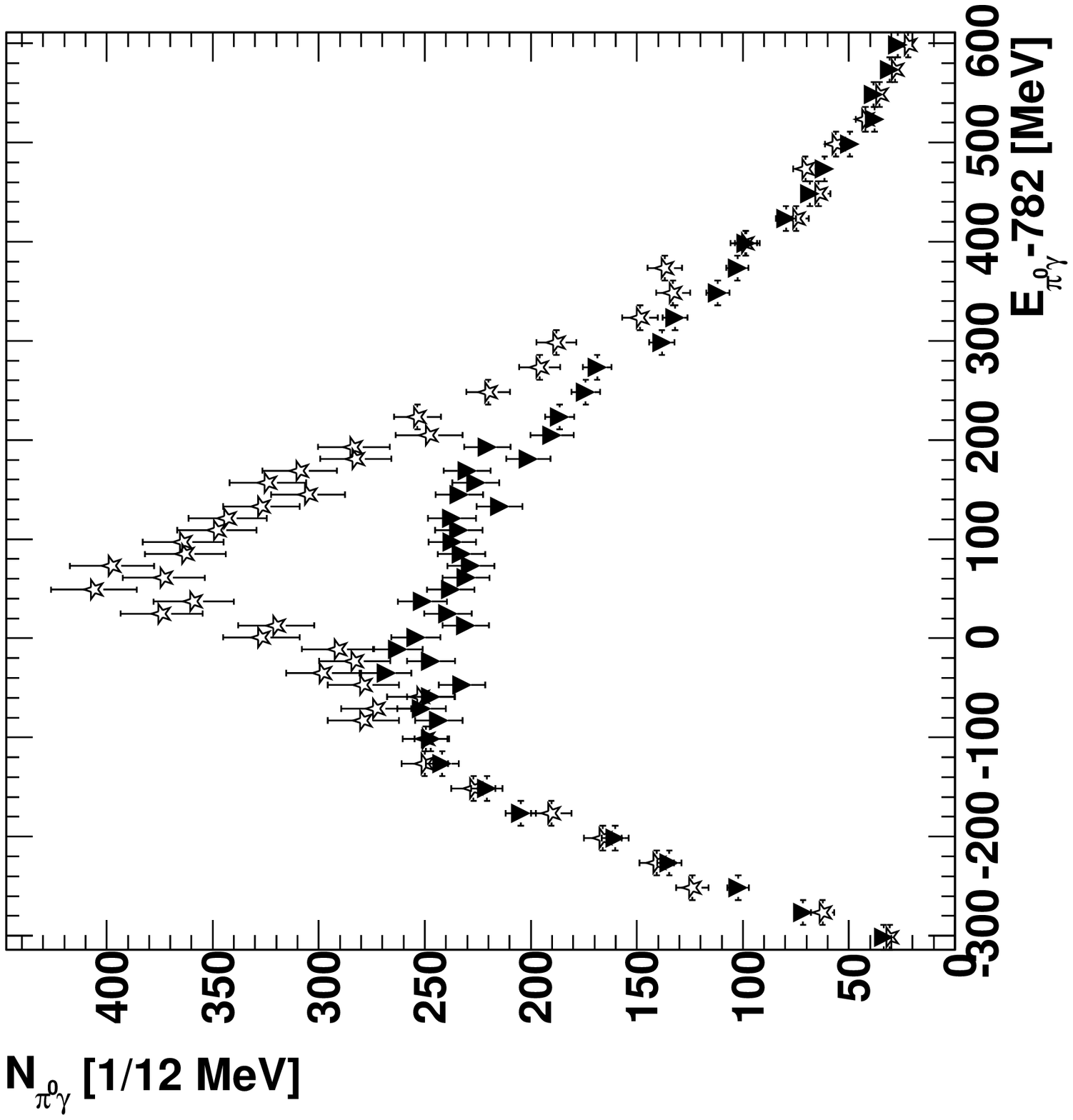} \includegraphics[height=0.91\textheight]{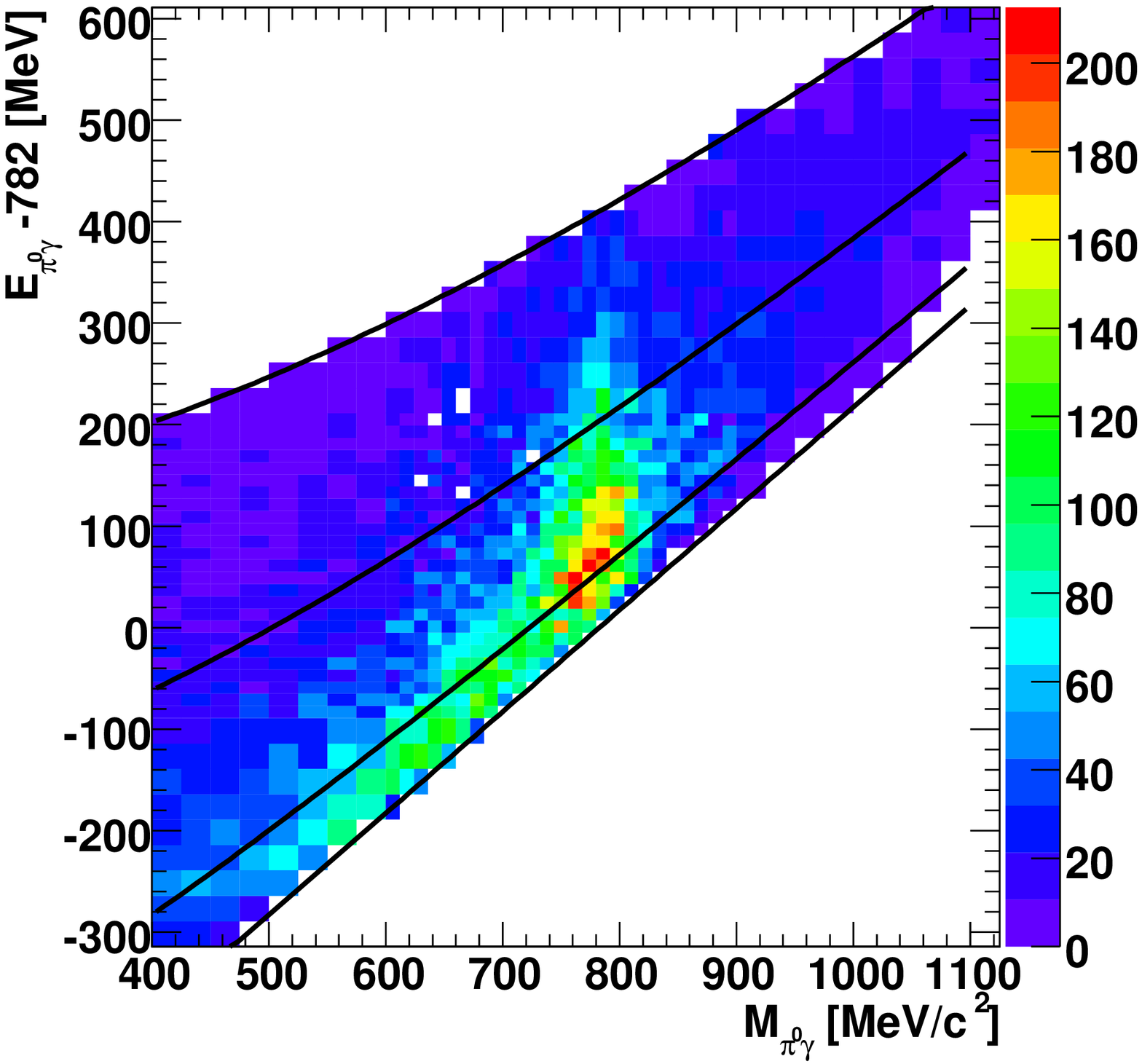}}
     \hspace*{6.3cm}{\includegraphics[height=0.28\textheight]{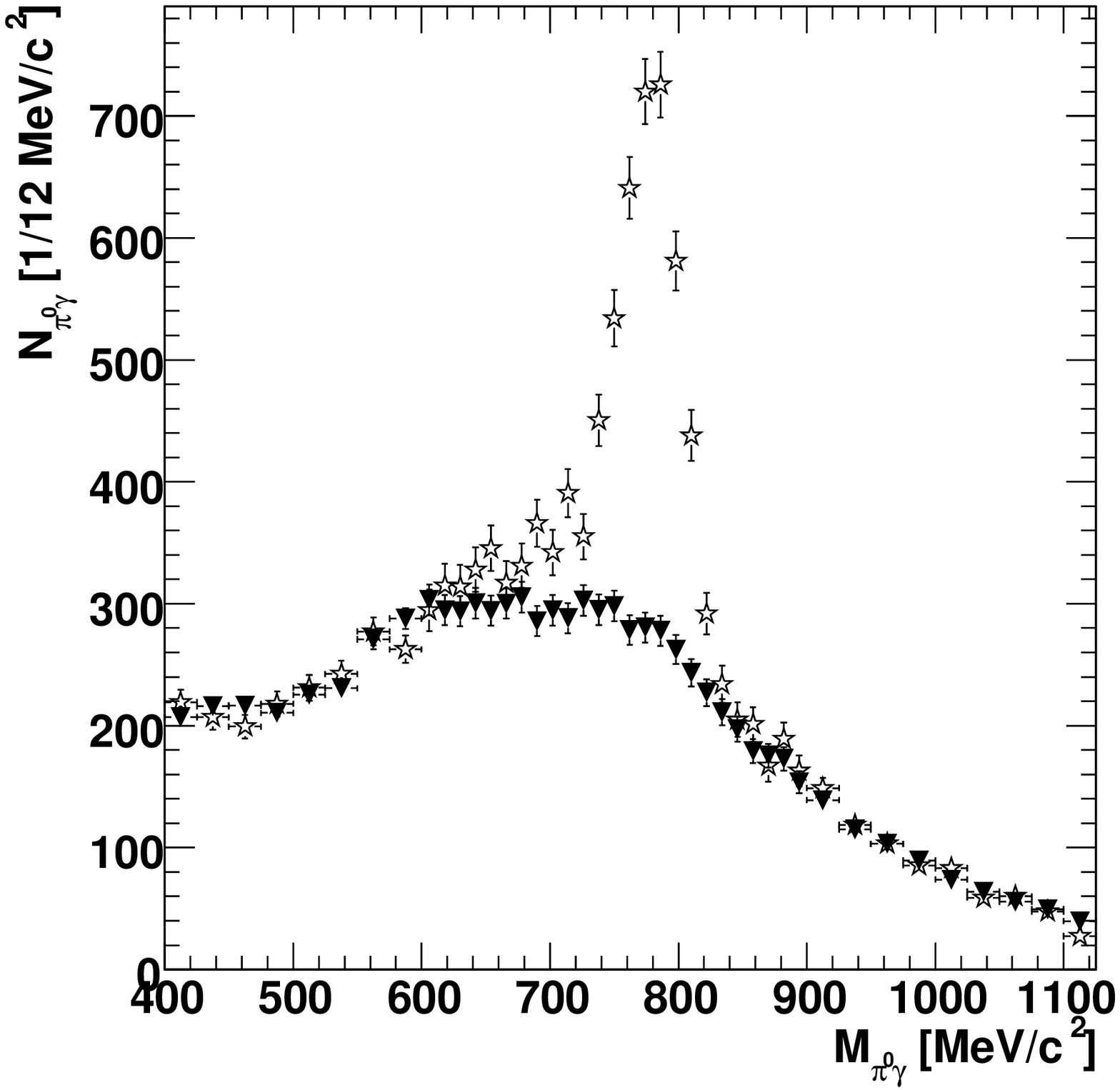}}  
\caption{Distribution of $\pi^0 \gamma$ events for the C target as a function of their invariant mass and the total energy of the $\pi^0 \gamma$ pairs minus the mass of the free $\omega$-meson (782 MeV/$c^2$). Black curves indicate lines of equal $\pi^0 \gamma$ pair momenta of 0, 300, 600, and 900 MeV/$c$. The spectra left and below are projections of the data onto the two axes, respectively. The data points (open stars) are compared to $\pi^0 \gamma$ background pairs (full triangles), arising from $\pi^0 \pi^0$ and $\pi^0 \eta$ events. See text for more details.}
\label{fig:plane}
\end{figure}

\begin{figure}[!ht]
 \resizebox{0.8\textwidth}{!}{
    \includegraphics[height=0.4\textheight]{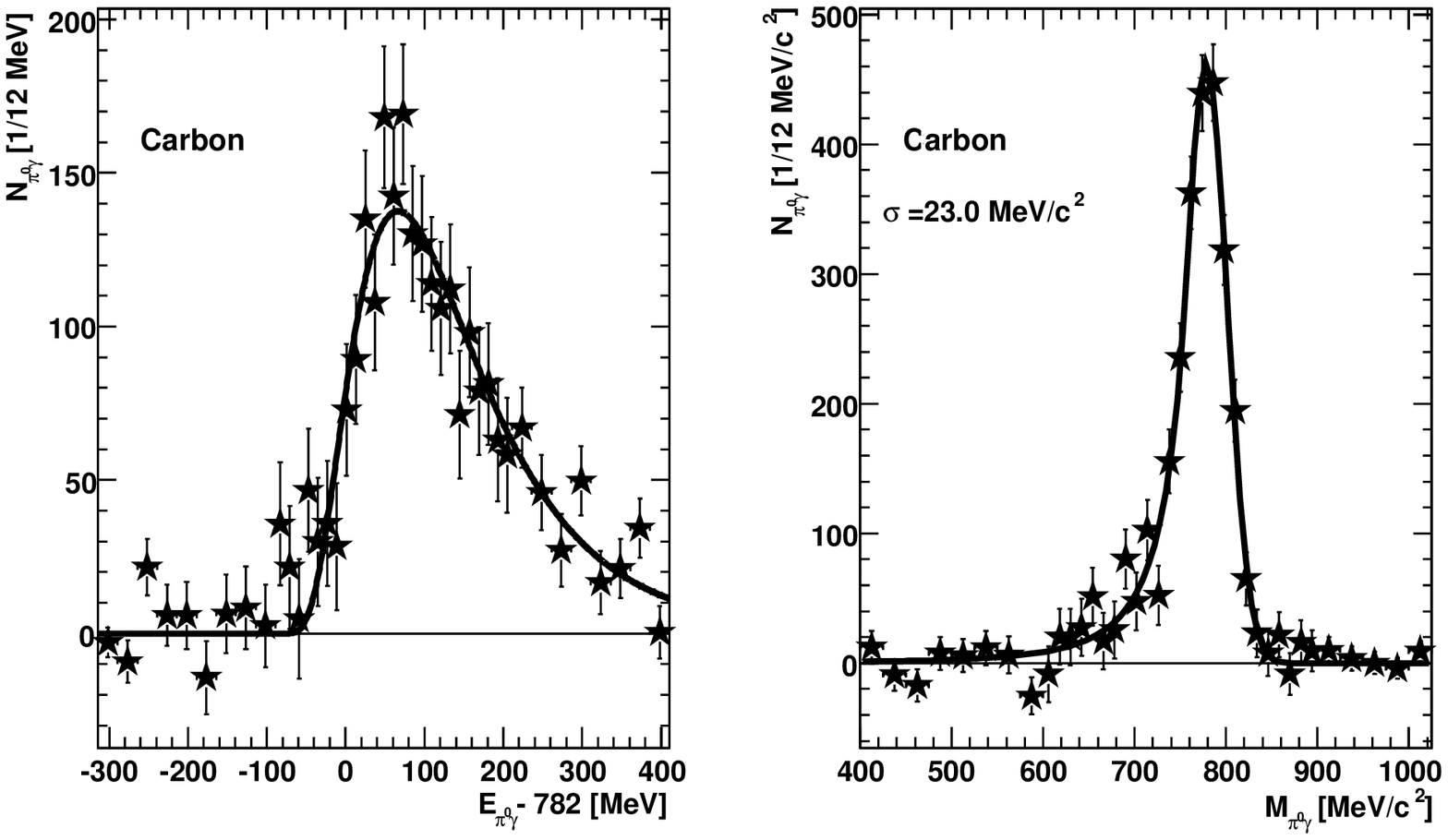}}
\caption{Distribution of the total energy of $\pi^0 \gamma$ pairs minus 782 MeV (left) and invariant mass distribution of $\pi^0 \gamma$ pairs after background subtraction for the C target (right). Only statistical errors are given. The solid curves represent fits to the energy and mass distributions with the Novosibirsk function \cite{Aubert} and the Crystal Ball detector response function  \cite{Gaiser}, respectively.}
\label{fig:projections_C}
\end{figure}

Fig.~\ref{fig:plane} shows the yield of reconstructed $\pi^0 \gamma$ pairs in a two-dimensional plot as a function of the $\pi^0 \gamma$ invariant mass and the total energy of the $\pi^0 \gamma$ pair minus the mass of the free $\omega$-meson of 782 MeV/$c^2$. Projections onto the respective axes exhibit clear signals above a background arising from $\pi^0 \pi^0$, $\pi^0 \eta \rightarrow 4 \gamma$ events where one of the 4 photons in the final state escaped undetected. For determining the background contributions from the $\pi^0 \pi^0$ and $\pi^0 \eta$ channels, events with 1 charged hit in MiniTAPS and 4 neutral hits elsewhere were analyzed. Applying the method developed in \cite{Nanova_PRC}, this background was directly determined from the data in an analysis of  the 4 photon events by arbitrarily omitting one of the 4 photons, irrespective of photon angle and energy, and otherwise applying the same cuts. The background shape is nicely reproduced by this analysis of 4 photon events. The background spectra were normalized to the background in the signal spectra outside of the signal region (400 MeV/$c^2 \le m_{\pi^0 \gamma} \le 600$ MeV/$c^2$ and 864 MeV/$c^2 \le m_{\pi^0 \gamma} \le 1200$ MeV/$c^2$), separately for the different momentum bins indicated in Fig.~\ref{fig:plane} (for carbon: 0 - 300 MeV/$c$, 300 - 600 MeV/$c$, 600 - 900 MeV/$c$). After subtracting this background 2500 events remained; their energy and invariant mass distributions are shown in Fig.~\ref{fig:projections_C}. Statistical errors were determined from the signal (S) and background (BG) yields according to the formula $\Delta S = \sqrt{(S+2BG)}$.\\

\begin{figure}[!ht]\

 \resizebox{0.5\textwidth}{!}{
    \includegraphics[height=0.4\textheight]{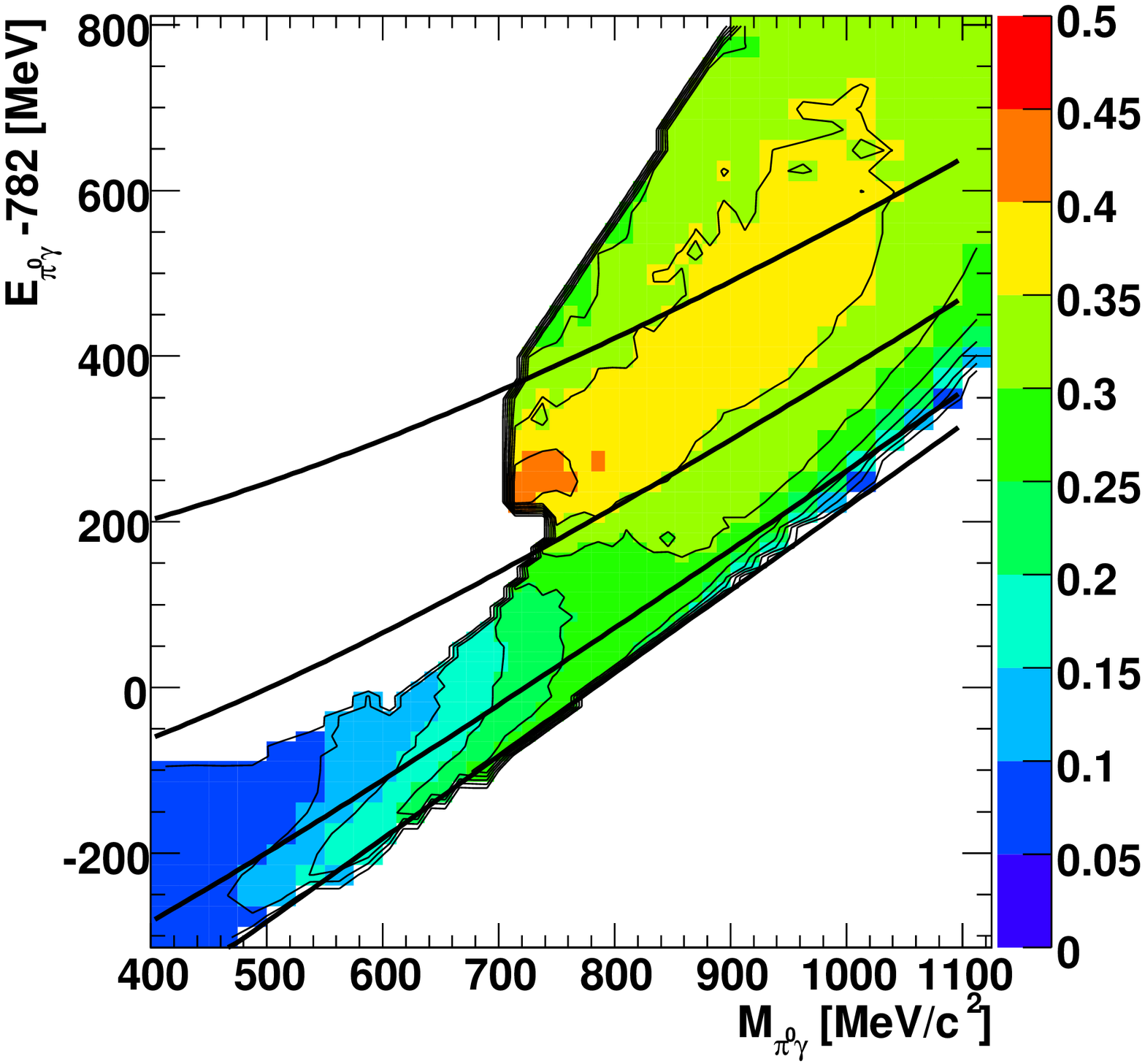}}
\caption{Acceptance for $\pi^0 \gamma$ pairs in coincidence with a proton in MiniTAPS ($\Theta_p = 1^{\circ} - 11^{\circ}$). The contour lines indicate increments by 10$\%$. Black curves indicate lines of equal $\pi^0 \gamma$ pair momenta of 0, 300, 600, and 900 MeV/$c$, as in Fig.~\ref{fig:plane}.}
\label{fig:acc}
\end{figure}

For converting the observed distributions into cross sections the acceptance of the detector system had to be determined for the given kinematics. Simulations were performed using the GEANT3 package with the complete detector setup implemented. In the event generator, $\pi^0 \gamma$ pairs with invariant masses between 400 to 1200 MeV/$c^2$ were produced in coincidence with protons going into the angular range of $1^{\circ} - 11^{\circ}$ covered by the MiniTAPS array for the incident photon energy range of 1250 - 3100 MeV. The simulated data were subjected to the same cuts as applied in the data analysis. Fig.~\ref{fig:acc} shows the resulting acceptance distribution for $\pi^0\gamma$ pairs under the constraint that a proton was identified in MiniTAPS by the proton band cut (Fig.~\ref{fig:p_band}).
The two-dimensional distributions of data and simulated acceptance are discretized in bins of 12x12 MeV for 600 MeV/$c^2 \le m_{\pi^0 \gamma} \le 900$ MeV/$c^2$ and -89 MeV $\le E_{\pi^0 \gamma}-782$ MeV$ \le 211$ MeV and in bins of 25x12 MeV elsewhere, except for the areas near the outer corners of the two dimensional plane of 
Fig.~\ref{fig:plane} where a coarser binning of 25x25 MeV was used. After a two dimensional subtraction of the background from the data (see Fig.~\ref{fig:plane}) the remaining signal counts are corrected for acceptance bin by bin. 

Systematic uncertainties in the cross section determinations arise from the background subtraction in the determination of the $\omega$ yield, mainly from the normalization of the 4 neutral hit events (10 - 15$\%$), from the acceptance simulation (5$\%$), the photon flux determination (5 - 10$\%$, see \cite{Nanova_potential}), and uncertainties in the photon initial state interaction (photon shadowing \cite {Bianchi}).
The contributions to the total systematic error are listed in Table I.
\begin{table}[!ht]
\centering
\caption{Sources of systematic errors}
\begin{footnotesize}
\begin{tabular}{ll}
\hline
background subtraction & $\approx$ 10 - 15$\%$\\
acceptance & $\lesssim 10\%$\\
photon flux & 5 - 10 $\%$\\
photon shadowing & 5$\%$\\
\hline
total& $\approx 20\%$\\
\end{tabular}
\end{footnotesize}
\label{tab:syst}
\end{table}

\section{Results and Discussions}

\begin{figure}[!ht]
 \resizebox{0.8\textwidth}{!}{
    \includegraphics[height=0.4\textheight]{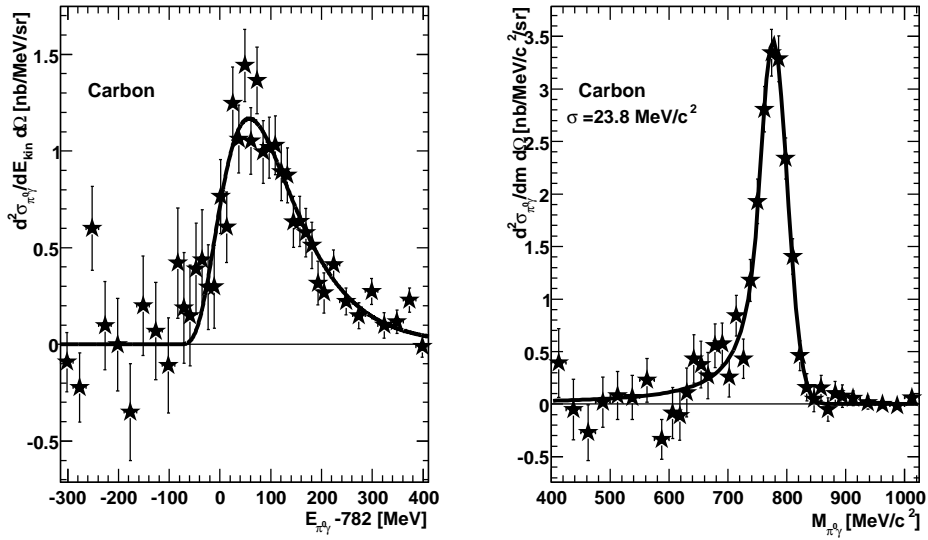}}
\caption{Left: Differential cross section for the photo production of $\omega$-mesons off C in coincidence with protons in $\Theta_p = 1^{\circ}-11^{\circ}$ as a function of the total energy of the $\pi^0 \gamma$ pairs minus 782 MeV. The data have been fitted with the Novosibirsk function \cite{Aubert}. Right: Differential cross section for  $\pi^0 \gamma$ pairs as a function of the  $\pi^0 \gamma$ invariant mass, fitted with the Crystal Ball detector response function \cite{Gaiser}.}
\label{fig:C_sigma}
\end{figure}

\begin{figure}[!ht]
 \resizebox{0.8\textwidth}{!}{
    \includegraphics[height=0.4\textheight]{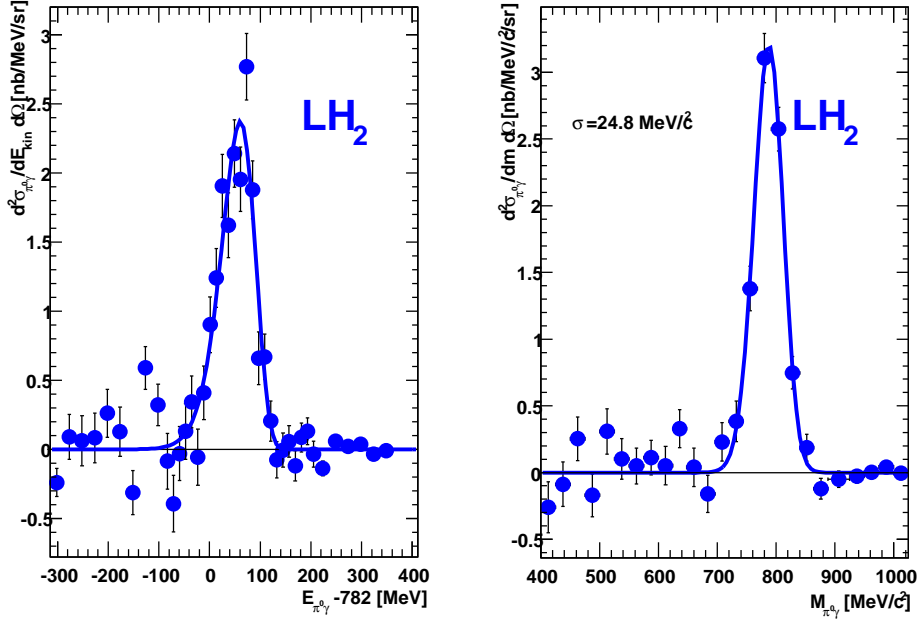}}
\caption{Left: Differential cross section for the photo production of $\omega$-mesons off the proton in coincidence with protons in $\Theta_p = 1^{\circ}-11^{\circ}$ as a function of the total energy of the $\pi^0 \gamma$ pairs minus 782 MeV. Right: Differential cross section for  $\pi^0 \gamma$ pairs as a function of the $\pi^0 \gamma$ invariant mass. The data have been fitted with the Novosibirsk function \cite{Aubert}.}
\label{fig:LH2_sigma}
\end{figure}

\begin{figure}[!ht]
 \resizebox{0.6\textwidth}{!}{
    \includegraphics[height=0.3\textheight]{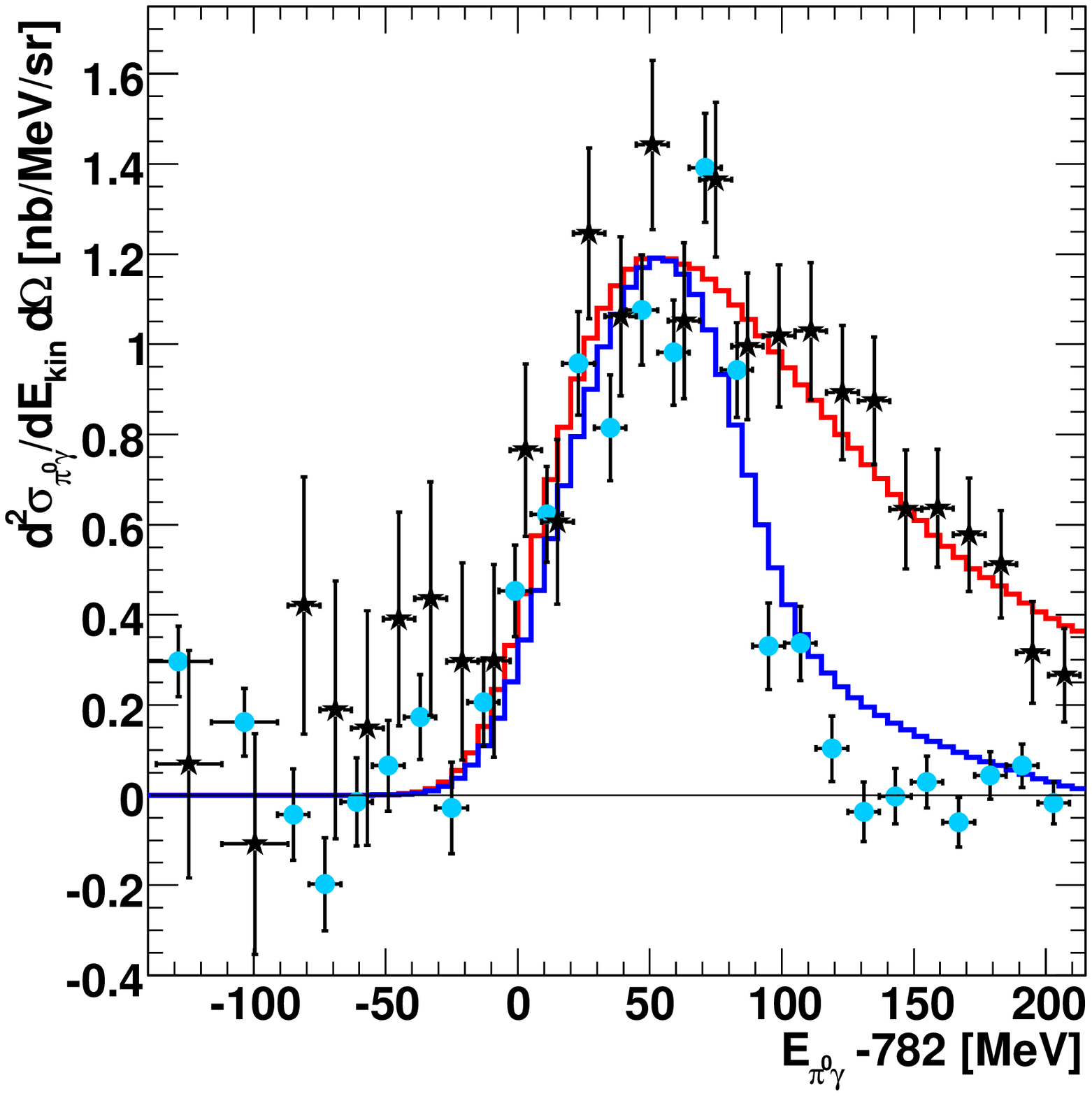}}
\caption{Comparison of the differential cross sections as a function of the total energy of $\pi^0 \gamma$ pairs minus 782 MeV for photo production of $\omega$-mesons off the proton (full blue circles) and carbon target (black stars), respectively. The data points are shifted by $\pm$ 2 MeV to avoid an overlap of the error bars for the two targets. The LH$_2$ data are normalized to the C data in the peak of the total energy distribution. The experimental distributions are compared to Monte Carlo simulations (LH$_2$ : blue histogram; C: red histogram), taking the Fermi motion of nucleons into account for the C target. All distributions request the detection of a proton in the polar angular range $1^{\circ}-11^{\circ}$ and are normalized to the fitted peak height for C in Fig.~\ref{fig:C_sigma} (left). The Monte Carlo simulations are folded with the experimental resolution of $\sigma_E\approx$ 16 MeV \cite{Aker}.}
\label{fig:E_LH2_C_Fermi}
\end{figure}

The cross sections obtained after acceptance corrections for the carbon target are shown in Fig.~\ref{fig:C_sigma} as a function of energy and invariant mass. By applying the same cuts, an analogous analysis of the data measured on the LH$_2$ target leads to the cross sections given in Fig.~\ref{fig:LH2_sigma}. A direct comparison of the energy distributions of $\pi^0 \gamma$ pairs measured on the C and LH$_2$ targets is shown in Fig.~\ref{fig:E_LH2_C_Fermi}.
The experimental data are compared to Monte Carlo simulations requesting the detection of a proton in MiniTAPS ($1^{\circ}-11^{\circ}$). For the carbon target the simulation takes the Fermi motion of nucleons into account which leads to the observed broadening of the energy distribution. It is remarkable that the distributions measured on the LH$_2$ and C targets peak at almost the same kinetic energy (C: 60.5$\pm$7 MeV; LH$_2$: 60.3$\pm$3.1 MeV), corresponding to $\omega$ momenta of about 300 MeV/$c$, about a factor two larger than the average momentum of nucleons bound in carbon. This is due to the fact that the forward going proton takes over most of the momentum of the incoming photon beam. Being so low in energy, the $\omega$-mesons are sensitive to the $\omega$ nucleus potential. In case of a strong repulsive (attractive) interaction one would expect the peak in the kinetic energy distribution to be shifted to higher (lower) energies for the C target. In the comparison of the kinetic energy distributions for the LH$_2$ and carbon targets one has to consider that for the latter the photo production of the $\omega$-meson occurs off a bound nucleon which sees a nuclear potential as well. The participant nucleon emerging from the nucleus is subject to this potential which is known to be momentum dependent \cite{Rudy}. Studies taking these effects into account are under way \cite{Paryev_priv}.

Both distributions fall off rather steeply towards negative energies. Structures at negative energies in the energy distribution of the $\pi^0 \gamma$ pairs, which would indicate the population and decay of bound $\omega~^{11}$B states, are not observed. The $\pi^0\gamma$ cross section in the energy range from - 100 to 0 MeV is on average $(0.3 \pm0.1)$ nb/ MeV/sr. By correcting for the effective branching ratio for in medium $\omega \rightarrow\pi^0 \gamma $ decays of about 1.5 $\%$ (see below) a population cross section of $(22\pm7)$ nb/MeV/sr was deduced which is of the order of magnitude expected for the formation of $\omega$ mesic states \cite{Marco_Weise,Nagahiro_omega}. The $\pi^0 \gamma$ invariant mass distribution exhibits some tailing towards lower $\omega$ masses for the carbon target which is not observed for the LH$_2$ target. This tailing appears to be correlated with the tailing in the total energy distribution discussed below.

\begin{figure}[!ht]
 \resizebox{0.90\textwidth}{!}{
    \includegraphics[height=0.99\textheight]{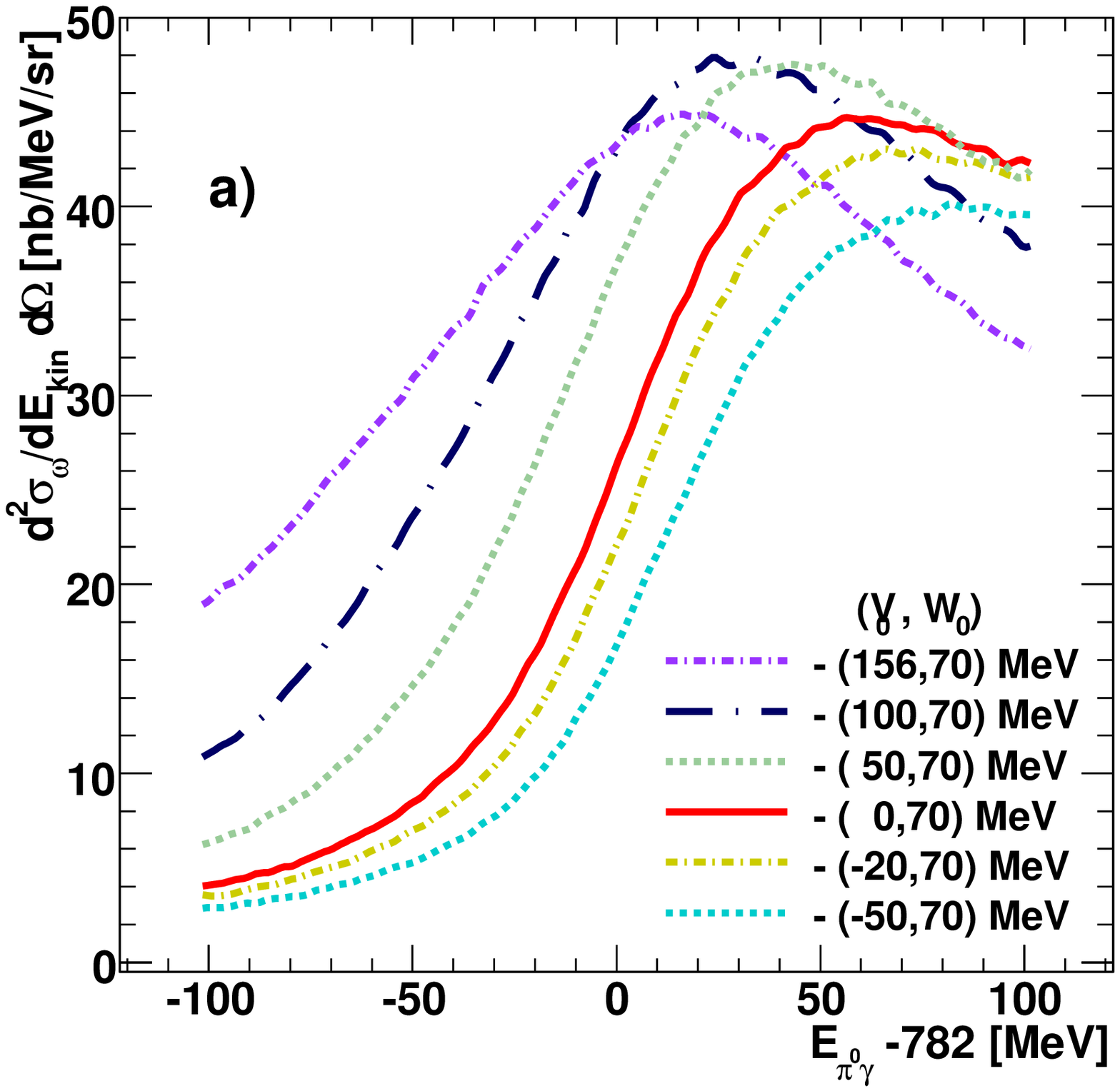} \includegraphics[height=0.99\textheight]{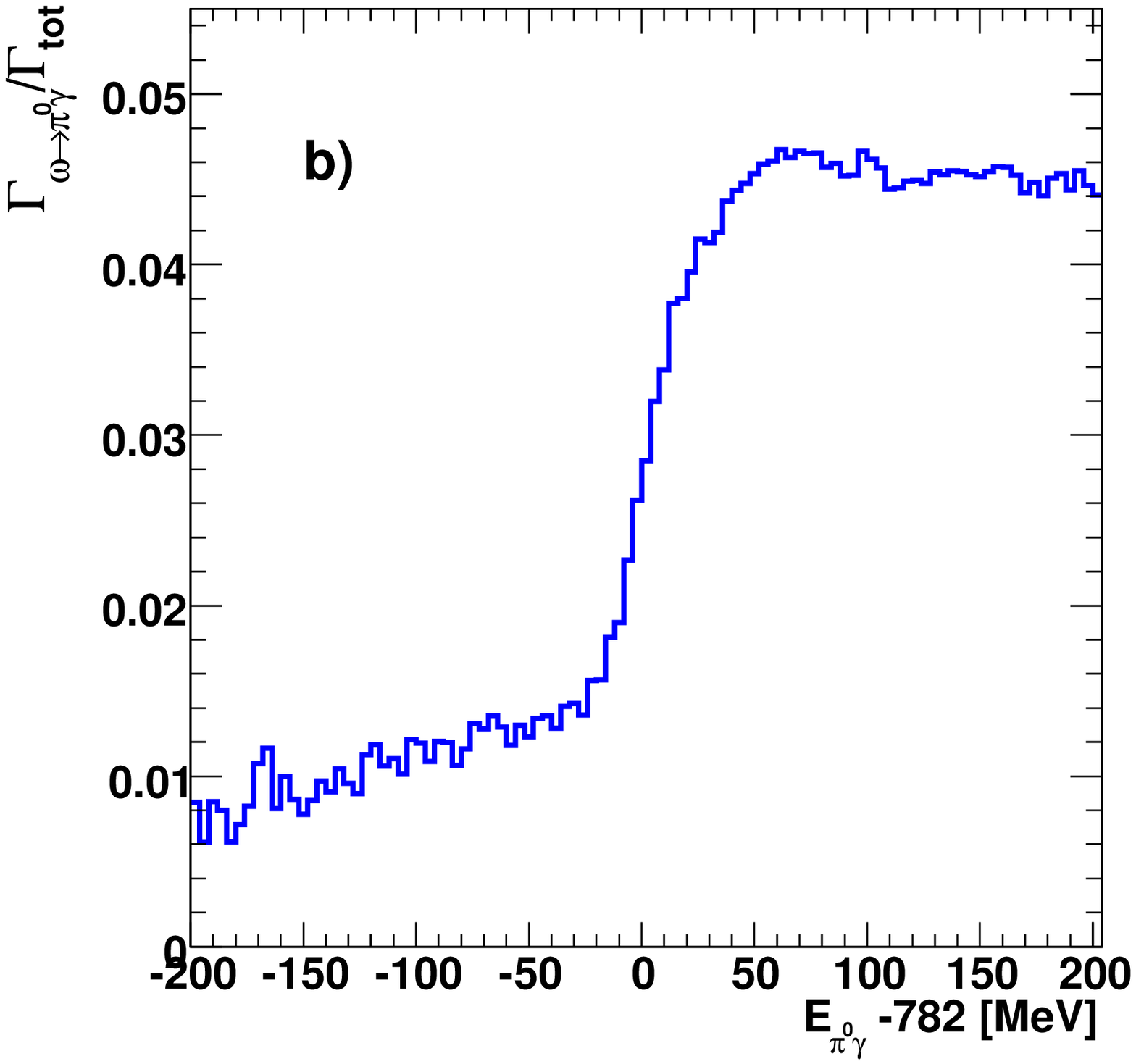}}
    {\includegraphics[height=0.28\textheight]{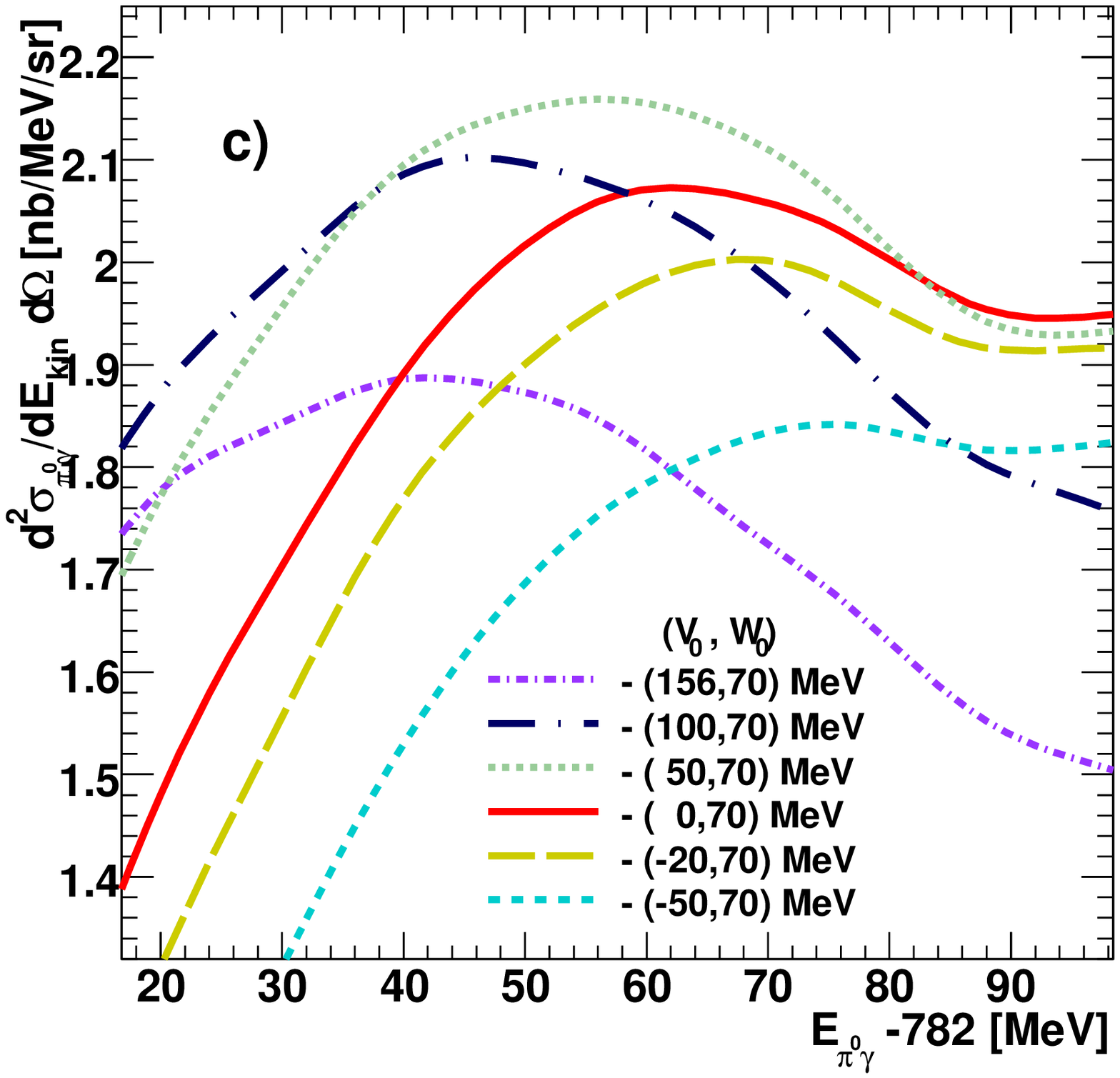} \includegraphics[height=0.28\textheight]{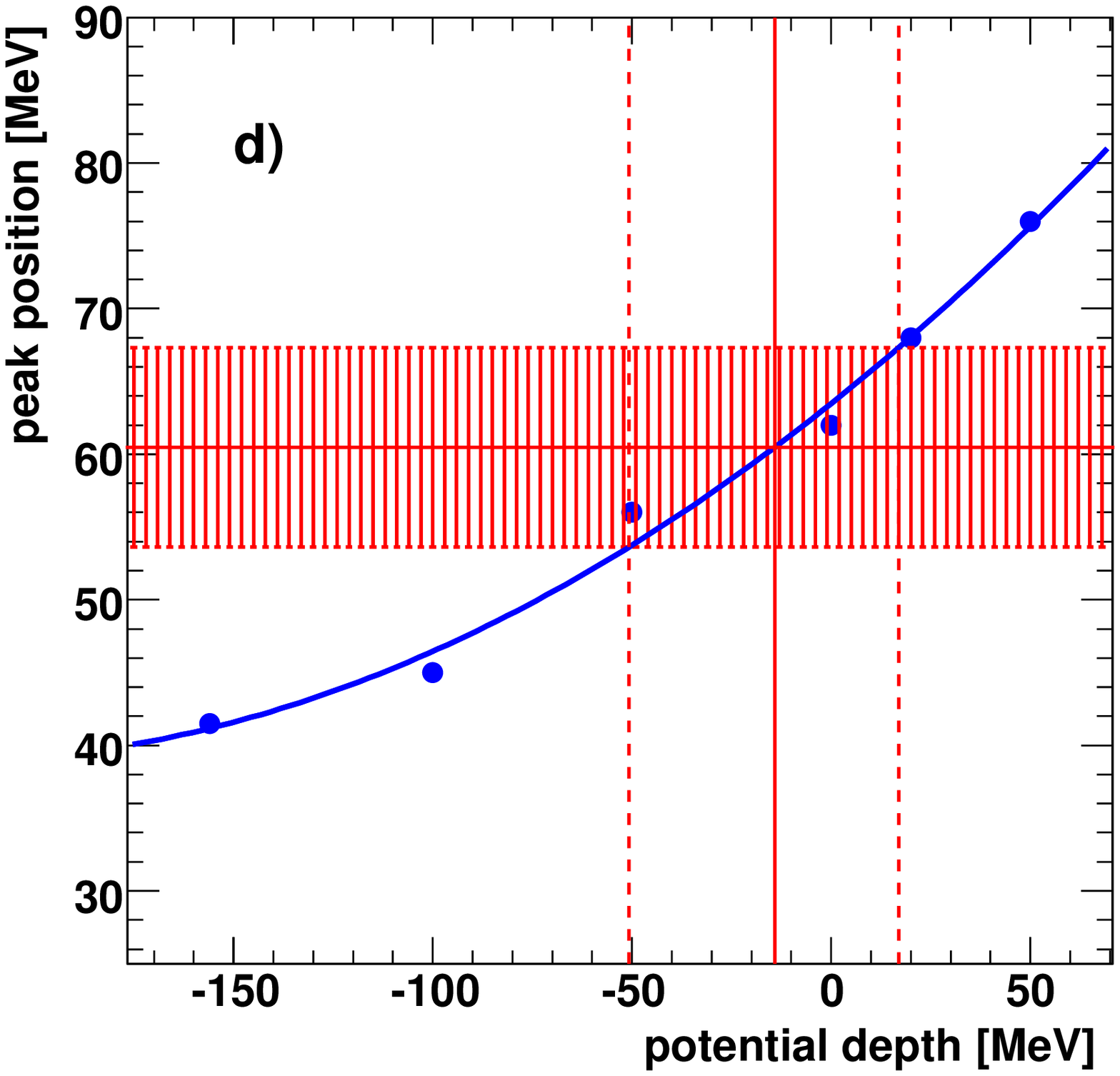}}
\caption{a): Differential cross section for $\omega$ production with the Green function method,  described in  \cite{Nagahiro_eta,Nagahiro_omega,Nagahiro_Zaki,Nagahiro_Kenta}, for different real potential depths $V_0$ and an imaginary potential of $W_0(\rho=\rho_0) $= -70 MeV after averaging over the incident photon energy interval of 1250 - 3100 MeV and the proton angular range of $1^{\circ} - 11^{\circ}$. b): Branching ratio for the decay of bound and free $\omega$-mesons into the $\pi^0 \gamma$ channel, deduced from GiBUU transport calculations (see text). c): $\pi^0 \gamma$ cross sections derived from the formation cross sections (a) by multiplying with the $\pi^0 \gamma$ branching ratio (b). Please note the changes in axes with respect to (a). d): Correlation between the potential depth and the peak position in the total energy distribution (c).  The (blue) points represent the peak positions in the total energy distribution for the different scenarios. The (blue) solid curve is a fit to the points. The red dashed area corresponds to the measured peak position of (60.5 $\pm$ 7) MeV (see Fig.~\ref{fig:C_sigma} left).}
\label{fig:formation_and_decay}
\end{figure}

For a more quantitative analysis the measured distributions were compared to calculations with the Green function method, described in detail in  \cite{Nagahiro_eta,Nagahiro_omega,Nagahiro_Zaki,Nagahiro_Kenta} which focus on the bound state region but also extend to 100 MeV above the production threshold. The calculations have been performed for an imaginary potential of $W_0(\rho=\rho_0)$= - 70 MeV, as determined by transparency ratio measurements \cite{Kotulla}, assuming different depths of the real potential. The cross sections for the formation of  $\omega ^{11}$B bound states have been calculated including the optical potential for the $\omega$ and the forward going proton. By averaging over the incident photon energy range of 1250 - 3100 MeV and the proton angle ($1^{\circ}-11^{\circ}$), the cross sections shown in Fig.~\ref{fig:formation_and_decay} (a) were obtained after folding with the energy resolution of $\sigma_E$ = 16 MeV.

For a comparison with the experimentally measured $\pi^0 \gamma$ cross section the formation cross sections were multiplied by the branching ratio for the decay of the $\omega$ mesic states or recoiling $\omega$-mesons into the $\pi^0 \gamma$ channel. This branching ratio can be extracted from GiBUU transport simulations \cite{Buss} by comparing the total energy distribution of $\pi^0 \gamma$ pairs in the final state with the total energy distribution of $\omega$-mesons at the time of the depopulation  of the $\omega$ mesic states by either $\omega $ absorption or decay. Being a semi-classical non-equilibrium description of hadronic processes, the GiBUU transport model cannot describe the population of quantum-mechanical bound states but it can calculate the absorption and decay of $\omega$-mesons which can not leave the nucleus and are thus bound to it since their total energy is $E_{\pi^0 \gamma}$-782 MeV$\le$ 0. The resulting branching ratio is shown in Fig.~\ref{fig:formation_and_decay} (b). For negative total energies this branching ratio is on the level of 1.0 - 1.5$\%$. This value is reduced relative to the branching ratio for $\omega \rightarrow \pi^0 \gamma$ decays in vacuum because of the strong absorption in the nuclear medium where the  $\omega$-mesons are confined. For positive total energies $\omega$-mesons can escape from the nucleus and decay in vacuum. For the kinetic energies considered here, the branching ratio does, however, not reach the value for the free $\omega \rightarrow \pi^0 \gamma$ decay of 8.3$\%$ since a fraction of $\omega$ decays still occurs in the nuclear medium in competition to $\omega$ absorption. The expected $\pi^0 \gamma$ cross sections obtained by multiplying bin-by-bin the formation cross sections with the energy-dependent branching ratio are shown in Fig.~\ref{fig:formation_and_decay} (c).  The sensitivity of the peak position in the kinetic energy distribution on the potential depth, exhibited in Fig.~\ref{fig:formation_and_decay} (c), can be exploited to deduce the depth of the real part of the $\omega$-nucleus potential. The correlation between the potential depth and the peak in the kinetic energy distribution is plotted in Fig.~\ref{fig:formation_and_decay} (d). A comparison with the experimentally determined peak position indicates a potential depth of -(15$\pm$35) MeV which appears to be too small to allow for the formation and population of $\omega$ mesic states, in particular in view of the large imaginary potential of -70 MeV.

The systematic error in this determination arises on the one hand from uncertainties in fitting the peak position of the experimental kinetic energy distribution which have been determined by changing the fit range and by using different fit functions (e.g., Novosibirsk \cite{Aubert}, $f(E)=a \cdot (E-E_1)^{\alpha} \cdot (E-E_2)^{\beta}$) and are of the order of $\pm$ 5 MeV. Another source of systematic errors is the uncertainty in extracting the energy dependent branching ratio from the GiBUU code which has been estimated from the variation of the branching ratio for different in-medium modification scenarios. Corresponding changes in the correlation between the potential depth and the peak position in the kinetic energy distribution of the $\omega$-mesons lead to changes in $V_0$ by $\pm$ 20 MeV, giving the final result: $V_0 (\rho=\rho_0) = -(15\pm35($stat$)\pm$20(syst)) MeV. This result confirms the conclusions drawn from a comparison of measured $\omega$ momentum distributions \cite{Micha} with GiBUU model calculations; theoretically predicted large in-medium mass shifts of the order of -100 to -150 MeV \cite{Marco_Weise,Brown_Rho,Hatsuda_Lee,Klingl1,Klingl2,QMC} are not supported by experimental observations. The observed yield at negative total energies can be reconciled with this conclusion and explained as arising from the large imaginary part of the optical potential. Due to the large in-medium broadening of the $\omega$-meson, intensity is shifted towards negative energies as illustrated in Fig.~\ref{fig:formation_and_decay} (c) e.g., for the case of ($V_0,W_0$) = (0,-70 MeV).

\section{Conclusions}
A search for $\omega$ mesic states has been performed in a photo nuclear reaction on a carbon target for incident photon energies between 1250 and 3100 MeV. $\pi^0 \gamma$ decays of free and bound $\omega$-mesons were registered in coincidence with forward going protons. Statistically significant structures in the total energy distribution of $\pi^0\gamma$ pairs, which would indicate the population and decay of $\omega$ mesic states, have not been observed. The average cross section measured in the bound state region between total energies of -100 to 0 MeV of $(0.3\pm0.1$) nb/MeV/sr, corresponding to a formation cross section of $(22\pm7$) nb/MeV/sr, may arise from yield leaking into the bound state region due to the large in-medium width of the $\omega$-meson. Exploiting the sensitivity of the peak in the $\omega$ kinetic energy distribution on the depth of the real part of the $\omega$ nucleus optical potential, a value for the potential depth of $V_0(\rho=\rho_0) = -(15\pm35$(stat)$\pm$20(syst)) MeV has been deduced, excluding earlier theoretical predictions of attractive potentials with a depth of 100 - 150 MeV. 

\section*{Acknowledgments}
We thank the scientific and technical staff at ELSA for their important contribution to the success of the experiment. Illuminating and very helpful discussions with St. Leupold, U. Mosel and E. Paryev are highly acknowledged. This work was supported financially by the {\it Deutsche Forschungsgemeinschaft}  through SFB/TR16. The Basel group acknowledges support from the {\it Schweizerischer Nationalfonds}.

\end{document}